\documentclass[twocolumn,prb,amsmath,amssymb,showpacs,superscriptaddress,floatfix]{revtex4}

\usepackage{graphicx}

 \DeclareMathOperator{\Real}{\mathrm{Re}}

\begin{document}

\title{Density of states in \textit{d}-wave superconductors of finite size}

 \author{Ya.~V.~Fominov}
 \email{fominov@landau.ac.ru}
 \affiliation{L.~D.~Landau Institute for Theoretical Physics RAS, 119334 Moscow, Russia}

 \author{A.~A.~Golubov}
 \email{a.golubov@tnw.utwente.nl}
 \affiliation{Faculty of Science and Technology, University of Twente, 7500 AE Enschede, The~Netherlands}

\date{30 December 2004}

\begin{abstract}
We consider the effect of the finite size in the \textit{ab}-plane on the surface density of states (DoS) in clean
\textit{d}-wave superconductors. In the bulk, the DoS is gapless along the nodal directions, while the presence of a
surface leads to formation of another type of the low-energy states, the midgap states with zero energy. We demonstrate
that finiteness of the superconductor in one of dimensions provides the energy gap for all directions of quasiparticle
motion except for $\theta=45^\circ$ ($\theta$ is the angle between the trajectory and the surface normal); then the
angle-averaged DoS behaves linearly at small energies. This result is valid unless the crystal is $0^\circ$- or
$45^\circ$-oriented ($\alpha\ne 0^\circ$ or $45^\circ$, where $\alpha$ is the angle between the \textit{a}-axis and the
surface normal). In the special case of $\alpha=0^\circ$, the spectrum is gapped for all trajectories $\theta$; the
angle-averaged DoS is also gapped. In the special case of $\alpha=45^\circ$, the spectrum is gapless for all
trajectories $\theta$; the angle-averaged DoS is then large at low energies. In all the cases, the angle-resolved DoS
consists of energy bands that are formed similarly to the Kronig--Penney model. The analytical results are confirmed by
a self-consistent numerical calculation.
\end{abstract}


% 74.78.-w Superconducting films and low-dimensional structures
% 74.78.Bz High-Tc films
% 74.78.Fk Multilayers, superlattices, heterostructures
%
% 74.45.+c Proximity effects; Andreev effect; SN and SNS junctions

\maketitle

\section{Introduction}

Among many possible types of unconventional superconductors,\cite{Mineev} the materials with the \textit{d}-wave
symmetry of the pair potential are most widely discussed. The \textit{d}-wave symmetry is established in the
quasi-two-dimensional high-temperature cuprate superconductors.\cite{Van_Harlingen,Tsuei} A characteristic property of
the \textit{d}-wave superconductivity is the gapless spectrum of quasiparticles. The pair potential is anisotropic, and
the gap vanishes along the nodal directions. Another source of low-energy quasiparticles is the surface, which leads to
forming the midgap states (MGS);\cite{Hu} this phenomenon is due to the change of the sign of the pair potential along a
trajectory upon reflection. Thus a \textit{d}-wave superconductor is a gapless superconductor with two types of
low-energy quasiparticles.

The \textit{d}-wave superconductors can be employed in novel types of logic elements, qubits;\cite{Ioffe,Zagoskin} there
is experimental progress in this direction.\cite{Il'ichev,Tzalenchuk} However, the low-energy quasiparticles introduce
decoherence in \textit{d}-wave qubits.\cite{FGK,Amin_decoherence} At the same time, the authors of
Ref.~\onlinecite{Ioffe} mention a possibility to suppress the low-energy quasiparticles due to the finite size of the
\textit{d}-wave banks.

In this paper, we study the influence of the finite size of a \textit{d}-wave superconductor on the low-energy density
of states (DoS) at the surface. There is only a limited number of results related to this issue. The angle-averaged
surface DoS was numerically studied by Nagato and Nagai for $45^\circ$-oriented superconductors.\cite{NN} There is also
a number of numerical results on the DoS in clean SN systems (where S is a conventional \textit{s}-wave superconductor
and N is a normal metal), which can be relevant to the nodal directions of finite-size \textit{d}-wave superconductors
due to similarity of the pair potential profile along quasiparticle trajectories in the two systems (see below for
details). The works by van~Gelder \cite{vanGelder} and Gallagher \cite{Gallagher} are the most relevant in this respect.
%Finally, we mention the paper by Shelankov and Ozana,\cite{Shelankov} who numerically studied a junction between two
%conventional superconductors; their results are relevant for the $45^\circ$ trajectory in the \textit{d}-wave system.
In Ref.~\onlinecite{Shelankov}, Shelankov and Ozana suggested a method to treat multiple-interface superconducting
systems and, as an application, numerically considered the DoS in a finite-size bilayer. Their results are relevant for
the $45^\circ$ trajectory in the \textit{d}-wave system. Finally, we mention an analytical result of
Ref.~\onlinecite{Fauchere}, where Fauch\`{e}re \textit{et al.} considered an SN system with repulsive interaction
between the electrons in the N layer. In our language, their result refers to the splitting of the MGS.

\begin{figure}
 \includegraphics[width=23mm]{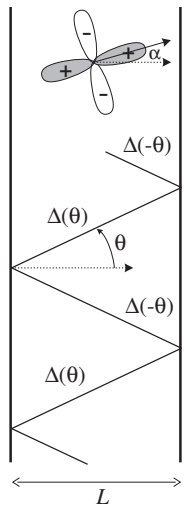}
\caption{\textit{d}$_{x^2-y^2}$-wave superconductor of finite width $L$ in the \textit{ab}-plane (quasi-two-dimensional
strip). The orientation of the crystalline \textit{a}-axis with respect to the surface normal is denoted by $\alpha$,
then the angular dependence of the pair potential is $\Delta(\theta) =\Delta_0 \cos( 2\theta -2\alpha)$. The pair
potential along the trajectory described by the angle $\theta$ changes periodically. We assume $0^\circ \leqslant \alpha
\leqslant 45^\circ$
--- this interval covers all physically different situations.}
 \label{fig:system}
\end{figure}

We demonstrate that the continuous normal-metallic spectrum along the nodal directions transforms to a set of energy
bands; the lowest band is separated from zero by a gap. The midgap states (existing in the finite intervals of the
trajectory angles $45^\circ -\alpha < |\theta| < 45^\circ +\alpha$ --- see Fig.~\ref{fig:system} for notations)
transform to two energy bands, situated symmetrically around zero, thus a gap appears. This gap depends on the direction
of the quasiparticle trajectory, and vanishes for exactly $\theta = 45^\circ$ direction (at any crystalline orientation
$\alpha$). After averaging over the directions, the DoS is zero at zero energy, and behaves linearly at small energies.
Thus the averaged surface DoS remains gapless but is strongly suppressed compared to that in the bulk \textit{d}-wave
superconductor (i.e., at $L\to \infty$).

In Sec.~\ref{sec:analytics}, we present analytical results for the DoS. In Sec.~\ref{sec:numerics}, we confirm and
illustrate these results by self-consistent numerical calculations. Finally, we present our conclusions in
Sec.~\ref{sec:conclusions}.

\section{Analytical results} \label{sec:analytics}

\begin{figure}
 \includegraphics[width=85mm]{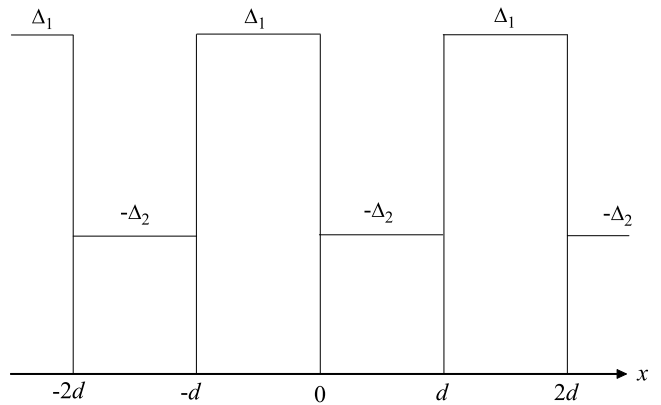}
\caption{The pair potential profile along a trajectory in the \textit{d}-wave superconductor of finite width. The signs
are chosen in such a way that the midgap states exist at $\Delta_1,\Delta_2 >0$. For definiteness, we choose $\Delta_1
\geqslant \Delta_2$; this does not impose any restrictions on the results. The pair potential changes on the surfaces of
the superconductor. With respect to Fig.~\ref{fig:system}, the introduced notations are defined as
$\Delta_1=\Delta(\theta)$, $\Delta_2=-\Delta(-\theta)$, and $d=L/\cos\theta$.}
 \label{fig:Delta12}
\end{figure}

We consider a quasi-two-dimensional \textit{d}$_{x^2-y^2}$-wave superconductor of finite width, i.e., a strip in the
\textit{ab}-plane
--- see Fig.~\ref{fig:system}. A quasiparticle trajectory is sequentially reflected from one or the other surface of the
strip, and the pair potential felt by the quasiparticle changes periodically. The profile of the pair potential along
the trajectory is schematically depicted in Fig.~\ref{fig:Delta12}. This profile is not self-consistent, while the
self-consistent pair potential is suppressed near the surfaces (see Sec.~\ref{sec:numerics} below). However, the width
of the regions where this happens has the characteristic scale of the coherence length $\xi = v_F / 2\pi T_c$ ($v_F$ is
the Fermi velocity, $T_c$ is the superconducting critical temperature\cite{T_c}). We assume $L \gg \xi$, then the
regions where $\Delta$ is suppressed are relatively narrow, hence the piecewise constant $\Delta$, depicted in
Fig.~\ref{fig:Delta12}, is a good approximation. The results of self-consistent numerical calculations will be discussed
below in Sec.~\ref{sec:numerics}; they agree with the analytical results of the present section.

The real-energy Eilenberger equations along the trajectory have the form (we choose $\Delta$ to be
real)\cite{Eilenberger}
\begin{gather}
\left( -2iE + v_F \frac\partial{\partial x} \right) f - 2\Delta g =0, \notag \\
\left( -2iE - v_F \frac\partial{\partial x} \right) \bar f - 2\Delta g =0, \label{Eilenberger} \\
v_F \frac\partial{\partial x} g +\Delta \left( \bar f -f \right) =0, \notag
\end{gather}
with the normalization condition
\begin{equation} \label{normalization}
g^2 + f \bar f =1.
\end{equation}
Here $g$ is the normal Green function, while $f$ and $\bar f$ are the anomalous Green functions that describe the
superconductivity.

Since the Green functions along the trajectory are continuous upon specular reflection at the surfaces of the
\textit{d}-wave superconductor, they obey the same condition in the effective one-dimensional problem described by
Fig.~\ref{fig:Delta12}: $g$, $f$, and $\bar f$ must be continuous at the edges of the intervals of constant $\Delta$.

The DoS (normalized by its normal-metal value) is determined by the real part of the normal Green function:
\begin{equation}
\nu = \Real g.
\end{equation}
The DoS is symmetric, $\nu(E)=\nu(-E)$.

In the bulk, the Green functions are
\begin{equation}
g = \frac{-iE}{\sqrt{\Delta^2-E^2}}, \qquad f = \bar f = \frac\Delta{\sqrt{\Delta^2-E^2}} ,
\end{equation}
which yields the standard DoS
\begin{equation}
\nu = \Real \frac{|E|}{\sqrt{E^2-\Delta^2}}.
\end{equation}

The general solution of the Eilenberger equations (\ref{Eilenberger}) is
\begin{gather}
\begin{pmatrix} g \\ f \\ \bar f \end{pmatrix} = A \begin{pmatrix} -iE \\ \Delta \\ \Delta \end{pmatrix} + \frac B2
\begin{pmatrix} \Delta \\ iE +\sqrt{\Delta^2-E^2} \\ iE-\sqrt{\Delta^2-E^2} \end{pmatrix} e^{kx} \notag \\
+ \frac C2 \begin{pmatrix} \Delta \\ iE -\sqrt{\Delta^2-E^2} \\ iE+\sqrt{\Delta^2-E^2} \end{pmatrix} e^{-kx},
\end{gather}
with
\begin{equation}
k=\frac{2\sqrt{\Delta^2-E^2}}{v_F}.
\end{equation}
The normalization condition (\ref{normalization}) yields:
\begin{equation} \label{normalization1}
A^2 + B C = \frac 1{\Delta^2-E^2}.
\end{equation}

When the pair potential is spatially symmetric, $\Delta(x)=\Delta(-x)$, the Green functions obey the relations
$g(x)=g(-x)$, $f(x)= \bar f(-x)$, hence $B=C$. In our case, the pair potential is symmetric with respect to the center
of each interval, thus we can write
\begin{widetext}
\begin{equation}
\begin{pmatrix} g \\ f \end{pmatrix} = A_1 \begin{pmatrix} -iE \\ \Delta_1 \end{pmatrix}
+ B_1 \begin{pmatrix} \Delta_1 \cosh( k_1 (x+d/2)) \\ iE \cosh(k_1 (x+d/2)) +\sqrt{\Delta_1^2-E^2} \sinh(k_1 (x+d/2))
\end{pmatrix}
\end{equation}
at $-d <x< 0$, and
\begin{equation}
\begin{pmatrix} g \\ f \end{pmatrix} = A_2 \begin{pmatrix} -iE \\ -\Delta_2 \end{pmatrix}
+ B_2
\begin{pmatrix} -\Delta_2 \cosh(k_2 (x-d/2)) \\ iE \cosh(k_2 (x-d/2)) +\sqrt{\Delta_2^2-E^2} \sinh(k_2 (x-d/2))
\end{pmatrix}
\end{equation}
at $0 <x< d$, while the solution on all the other intervals is obtained due to $2d$-periodicity of the Green functions.

The boundary conditions (continuity of the Green functions) at $x=0$ and $x=d$ yield four equations, only three of which
are independent. Solving this system of equations and using the normalization condition (\ref{normalization1}), we
determine all the four coefficients $A_1$, $A_2$, $B_1$, and $B_2$. Then we find the Green functions and the DoS at the
interface between the two intervals (this corresponds to the surface DoS in the \textit{d}-wave superconductor):
\begin{equation} \label{gd2}
\nu\left( x= 0 \right) = \Real \frac{E \left[ \sqrt{\Delta_1^2-E^2} \tanh\left(\frac{k_2 d}2 \right)
+\sqrt{\Delta_2^2-E^2} \tanh\left( \frac{k_1 d}2 \right) \right]}{\sqrt{ \left( \frac{\sqrt{\Delta_1^2-E^2}
\sqrt{\Delta_2^2-E^2}}{\cosh(k_1 d/2) \cosh(k_2 d/2)} \right)^2 - \left[ (E^2+\Delta_1 \Delta_2) \tanh\left(\frac{k_1
d}2\right) \tanh\left(\frac{k_2 d}2 \right) - \sqrt{\Delta_1^2-E^2} \sqrt{\Delta_2^2-E^2} \right]^2 }}.
\end{equation}
\end{widetext}
A similar formula was obtained by Gallagher;\cite{Gallagher,remark} at the same time, our results are different since
his formula refers to the center of the strip in the \textit{d}-wave problem, while we study the surface DoS.

The zero-energy DoS can be found immediately. If $\Delta_1 \ne \Delta_2$, then Eq. (\ref{gd2}) yields $\nu(E=0)=0$. If
$\Delta_1=\Delta_2$ ($\equiv \Delta$), then we obtain
\begin{equation}
\nu(E=0) = \cosh\left( \frac{\Delta d}{v_F} \right).
\end{equation}

In the above calculations, we did not assume $\Delta d/v_F \gg 1$, however it is necessary to assume this in order to
make the results physically sound, because in a \textit{d}-wave superconductor, the piecewise constant pair potential
(see Fig.~\ref{fig:Delta12}) is a good approximation only under this condition.

Below we analyze Eq. (\ref{gd2}) at low energies, $E \ll \Delta$, in the following relevant cases: a)~$\Delta_2=0$
(nodal directions), b)~$\Delta_1 \ne \Delta_2$, and c)~$\Delta_1=\Delta_2$ (the latter two cases correspond to the
presence of the MGS in the infinite system).

\subsection{Effect of the finite size on the nodal quasiparticles} \label{sec:nodal}

Let us consider a nodal direction, which corresponds to $\Delta_2 =0$. For brevity, we shall denote $\Delta_1$ by
$\Delta$.

In the bulk, the DoS along a nodal direction is normal-metallic,
\begin{equation}
\nu_{d\to\infty}=1.
\end{equation}

The finite-size problem was considered previously (although in a different context) by van~Gelder \cite{vanGelder} and
Gallagher.\cite{Gallagher} While van Gelder considered an academic one-dimensional model with periodic step function
$\Delta(x)$, in the work by Gallagher this situation emerged along a quasiparticle trajectory in a
superconductor--normal-metal sandwich (with conventional \textit{s}-wave superconductor). They numerically demonstrated
that in this superconducting version of the Kronig--Penney model,\cite{KronigPenney} the quasiparticle spectrum consists
of energy bands with square-root singularities at the band edges. Below we present analytical results for this problem.

Taking into account that
\begin{gather}
\frac{\Delta d}{v_F} \gg 1,\\
E \ll \Delta,
\end{gather}
we obtain from Eq. (\ref{gd2}):
\begin{widetext}
\begin{equation}
\nu = \Real \frac{\Delta \sin\left( \frac{E d}{v_F} \right) + E\cos\left( \frac{E d}{v_F} \right)} {\sqrt{ - \left[ E
\sin\left( \frac{E d}{v_F} \right) - \Delta \cos\left( \frac{E d}{v_F} \right) + 2\Delta \exp\left( -\frac{\Delta
d}{v_F} \right) \right] \left[ E \sin\left( \frac{E d}{v_F} \right) - \Delta \cos\left( \frac{E d}{v_F} \right) -2\Delta
\exp\left( -\frac{\Delta d}{v_F} \right) \right] }}.
\end{equation}
The bands are situated around the energies at which $E \sin\left( E d /v_F \right) - \Delta \cos\left( E d /v_F
\right)=0$. Since we consider $E\ll \Delta$, then $\cos\left( E d /v_F \right)$ must be very small to fulfill this
equation, hence its argument is very close to $\pi/2$ (for the lowest band). Thus the center of the lowest band is
\begin{equation} \label{E_0}
E_0 = \frac \pi 2 \frac{v_F}d \left( 1-\frac{v_F}{\Delta d} \right) \approx  \frac \pi 2 \frac{v_F}d .
\end{equation}
In the vicinity of $E_0$, the DoS can be written as
\begin{equation}
\nu = \Real \frac{v_F /d} {\sqrt{ - \left[ E -E_0 +2 \frac{v_F}d \exp\left( -\frac{\Delta d}{v_F} \right) \right] \left[
E -E_0 -2 \frac{v_F}d \exp\left( -\frac{\Delta d}{v_F} \right) \right] }},
\end{equation}
\end{widetext}
hence the width of the band is
\begin{equation} \label{deltaE}
\delta E = 4 \frac{v_F}d \exp\left( -\frac{\Delta d}{v_F} \right),
\end{equation}
and the DoS has integrable square-root singularities at the edges of the band.

The minimal value of the DoS in the band is achieved at $E = E_0$, and equals
\begin{equation}
\nu_\mathrm{min} = \frac 12 \exp\left( \frac{\Delta d}{v_F} \right).
\end{equation}

The physical mechanisms behind the above results are quite transparent. Instead of the normal-metallic situation that
takes place for the nodal directions in the bulk, in the finite system we obtain the profile of the pair potential
corresponding to the SN superlattice (Fig.~\ref{fig:Delta12} with $\Delta_2=0$). Then the energy spectrum in each normal
layer consists of the Andreev levels\cite{Andreev} which are smeared into the bands due to periodicity of the system.
The energy of the lowest Andreev level corresponds to the center of the band, see Eq. (\ref{E_0}), while smearing is due
to tunneling across the barrier of height $\Delta$ and width $d$, and thus contains the tunneling exponential, see Eq.
(\ref{deltaE}).

\subsection{Effect of the finite size on the midgap states}

In the infinite system ($L,d \to\infty$), the midgap states arise if the pair potential changes its sing upon reflection
from the surface.\cite{Hu} According to our definitions (Figs.~\ref{fig:system} and~\ref{fig:Delta12}), this happens at
$\Delta_1,\Delta_2 >0$. The MGS are localized near the surfaces and have exactly zero energy; the corresponding DoS is
\begin{equation}
\nu_{d\to \infty} = 2\pi \frac{\Delta_1 \Delta_2}{\Delta_1 +\Delta_2} \delta(E).
\end{equation}
Below we consider the effect of finite $d$ on this result; the results for the cases of differing and coinciding
$\Delta_1$ and $\Delta_2$ will be qualitatively different.

\subsubsection{The case $\Delta_1>\Delta_2$} \label{sec:Delta1greater2}

Let us consider the general case, when $\Delta_1$ and $\Delta_2$ are nonzero and can be different (still, the sings are
such that the MGS exist in the infinite system). Employing
\begin{gather}
\frac{\Delta_1 d}{v_F}, \frac{\Delta_2 d}{v_F} \gg 1,\\
E \ll \Delta_1, \Delta_2,
\end{gather}
and expanding Eq. (\ref{gd2}), we obtain:
\begin{widetext}
\begin{gather}
\nu = \frac{2\Delta_1 \Delta_2}{\Delta_1+\Delta_2} \Real \frac E{\sqrt{ - \left[ E^2 - \left( \frac{2\Delta_1
\Delta_2}{\Delta_1+\Delta_2} \right)^2 (e_2-e_1)^2 \right] \left[ E^2 - \left( \frac{2\Delta_1
\Delta_2}{\Delta_1+\Delta_2} \right)^2 (e_2+e_1)^2 \right] }}, \label{nu} \\
e_1 = \exp\left( -\frac{\Delta_1 d}{v_F} \right),\qquad e_2 = \exp\left( -\frac{\Delta_2 d}{v_F} \right). \notag
\end{gather}
\end{widetext}
This yields two bands, situated symmetrically around zero energy. The edges of the positive-energy band are
\begin{gather}
E_{b,\mathrm{min}} = \frac{2\Delta_1 \Delta_2}{\Delta_1+\Delta_2} \left[ \exp \left( - \frac{\Delta_2
d}{v_F} \right) - \exp \left( -\frac{\Delta_1 d}{v_F} \right) \right], \label{Emin} \\
E_{b,\mathrm{max}} = \frac{2\Delta_1 \Delta_2}{\Delta_1+\Delta_2} \left[ \exp \left( - \frac{\Delta_2 d}{v_F} \right) +
\exp \left( -\frac{\Delta_1 d}{v_F} \right) \right]. \label{Emax}
\end{gather}
The minimal value of the DoS in the band is achieved at $E = \frac{2\Delta_1 \Delta_2}{\Delta_1+\Delta_2} \sqrt{e_2^2
+e_1^2}$, and equals
\begin{equation}
\nu_\mathrm{min} =\frac 12 \sqrt{ \frac 1{e_1^2} +\frac 1{e_2^2} } \sim \frac 12 \exp\left( \frac{\Delta_1 d}{v_F}
\right).
\end{equation}
At the edges of the bands, the DoS has integrable square-root singularities.

The position of the center of the band in the limit $\Delta_1 \gg \Delta_2$ was calculated in Ref.~\onlinecite{Fauchere}
(although in a different context), while only discrete energy levels were discussed and the width of the band was not
studied.

Physically, the obtained results can be explained as follows. In the infinite system, $d\to \infty$, we have the
zero-energy levels localized near the interfaces between $\Delta_1$ and $-\Delta_2$. The first effect of the finite $d$
is to split the levels at the two neighboring interfaces due to tunneling across the $\Delta_2$ barrier (the lowest
barrier). This splitting is symmetric with respect to zero energy. Physically, it is similar to the standard
quantum-mechanical problem of the level splitting in the double-well potential.\cite{LL} The second effect of the finite
$d$ is to smear each of the split levels due to periodicity of the system; the smearing is due to tunneling across the
$\Delta_1$ barriers. This is similar to the standard Kronig--Penney model.\cite{KronigPenney} As a result, the center of
the band (\ref{Emin})--(\ref{Emax}) is determined by the tunneling exponential containing $\Delta_2$, while the width of
the band is determined by the tunneling exponential containing $\Delta_1$. Since $\Delta_2<\Delta_1$, the splitting of
the zero-energy level is larger than its smearing, hence a gap in the spectrum arises.

\subsubsection{The case $\Delta_1=\Delta_2$} \label{sec:Delta1=2}

At $\Delta_1=\Delta_2$ ($\equiv \Delta$), the DoS (\ref{nu}) reduces to
\begin{equation}
\nu = \frac\Delta{\sqrt{E_b^2 -E^2}},\qquad E_b =2\Delta \exp\left( -\frac{\Delta d}{v_F} \right).
\end{equation}
Thus the MGS is smeared into the band of width $2E_b$ around zero. The minimal value of the DoS is achieved at $E=0$,
and equals
\begin{equation} \label{nu_min}
\nu_\mathrm{min} =\frac 12 \exp\left( \frac{\Delta d}{v_F} \right),
\end{equation}
while at the edges of the band (at $E=\pm E_b$), the DoS has integrable square-root singularities.

This result means that the two bands (positive and negative) that existed at $\Delta_1 > \Delta_2$, touch each other at
$E=0$ and merge into a single band, while the singularities at $E=0$ transform into the minimum (\ref{nu_min}). The
equivalent result about the band centered at $E=0$ was numerically obtained in Ref.~\onlinecite{Shelankov} (although in
a different context).

The physical explanation of these results is the same as in the previous case, $\Delta_1>\Delta_2$. The only difference
is that in the case of equal barriers, $\Delta_1=\Delta_2$, the splitting of the zero-energy level is exactly the same
as its smearing, hence no gap in the spectrum appears.

\subsection{Angle-averaged DoS}

Let us consider the behavior of the angle-averaged surface DoS at $E\to 0$. The only contribution to the DoS arises from
the vicinity of the angles at which $\Delta_1=\Delta_2$ --- these are always the $\theta=\pm 45^\circ$ angles [at any
orientation $\alpha$, since $\Delta(\theta) =\Delta_0 \cos(2\theta-2\alpha)$].

Introducing $\vartheta$ via $\theta = \pi/4 -\vartheta$, we expand the pair potential in the vicinity of
$\theta=45^\circ$:
\begin{equation}
\Delta_1 \approx \Delta +\Delta' \vartheta, \qquad \Delta_2 \approx \Delta -\Delta' \vartheta,
\end{equation}
where
\begin{equation}
\Delta=\Delta_0 \sin 2\alpha,\qquad \Delta' = 2\Delta_0 \cos 2\alpha.
\end{equation}
Then we expand
\begin{equation}
e_2-e_1 \approx \exp \left( -\frac{\Delta d}{v_F} \right) \frac{2\Delta' \vartheta d}{v_F}
\end{equation}
--- we assume that
\begin{equation} \label{condition}
\frac{\Delta' \vartheta d}{v_F} \ll 1
\end{equation}
(since $\Delta' \sim \Delta$, this condition implies that $\vartheta \ll v_F / \Delta d \ll 1$). The angle-resolved DoS
(\ref{nu}) then takes the form
\begin{widetext}
\begin{equation} \label{nu_ar}
\nu = \Delta\, \Real \frac E{\sqrt{ - \left\{ E^2 - \left[ \frac{2\Delta \Delta' \vartheta d}{v_F} \exp\left(
-\frac{\Delta d}{v_F} \right) \right]^2 \right\} \left\{ E^2 - \left[ 2\Delta \exp\left( -\frac{\Delta d}{v_F} \right)
\right]^2 \right\} }}.
\end{equation}
\end{widetext}
The contribution to the angle-averaged DoS comes from the interval $-\vartheta_b <\vartheta <\vartheta_b$, where
\begin{equation}
\vartheta_b =\frac E {\frac{2\Delta\Delta' d}{v_F} \exp\left( -\frac{\Delta d}{v_F} \right)}.
\end{equation}
Condition (\ref{condition}) for all $\vartheta$ up to $\vartheta_b$ is equivalent to
\begin{equation}
E \ll E_b,
\end{equation}
where $E_b = 2\Delta \exp\left( -\Delta d /v_F \right)$ is the upper edge of the band.

The angle-resolved DoS (\ref{nu_ar}) can be written as
\begin{equation}
\nu = \frac 1{\frac{4\Delta\Delta' d}{v_F} \exp\left( -\frac{2\Delta d}{v_F} \right)} \Real \frac E{\sqrt{
\vartheta_b^2(E) -\vartheta^2}}.
\end{equation}
The angle-averaged DoS is
\begin{equation} \label{nu_av}
\nu_\mathrm{av} = \frac 2\pi \int_{-\vartheta_b}^{\vartheta_b} \nu(\vartheta) d\vartheta = \frac E{\frac{2\Delta\Delta'
d}{v_F} \exp\left( -\frac{2\Delta d}{v_F} \right)}.
\end{equation}
The DoS is zero at zero energy, and behaves linearly at small $E$. Thus the averaged surface DoS remains gapless but is
strongly suppressed compared to that in the bulk \textit{d}-wave superconductor (i.e., at $L\to \infty$).

The gapless structure of the angle-averaged DoS (\ref{nu_av}) is due to integrating in the vicinity of $\theta=\pm
45^\circ$, where the gap in the angle-resolved DoS vanishes. At the same time, the DoS probed by transport
methods\cite{TK} is given by
\begin{equation}
\nu_\mathrm{tr} \sim \int \nu(\theta) D(\theta) d\theta,
\end{equation}
and differs from Eq. (\ref{nu_av}) by the weighting factor $D(\theta)$, the angle-dependent transparency of the
tunneling interface. When the tunneling interface has finite thickness, the function $D(\theta)$ is exponentially
suppressed at not too small $\theta$, then the contribution of the $\theta = \pm 45^\circ$ trajectories can be
significantly suppressed [since $\nu_\mathrm{tr} \sim D(45^\circ) \nu_\mathrm{av}$ at low energies]. In this case, we
can expect that the transport DoS $\nu_\mathrm{tr}$ will be gapped despite the angular averaging. The directional
selectivity of tunneling is most pronounced in the scanning tunneling spectroscopy experiments, where the effective
tunneling cone around the surface normal can be as narrow as $\delta\theta \sim 20^\circ$.\cite{Leibovitch}

The result (\ref{nu_av}) for the angle-averaged DoS does not refer to the cases $\alpha=0^\circ$ and $\alpha= 45^\circ$
(and close orientations); these cases are special.

At $\alpha=0^\circ$, the MGS do not appear, and the low-energy DoS is entirely due to the nodal directions. Then
according to Sec.~\ref{sec:nodal}, the spectrum along a nodal direction acquires a gap due to the finite size. In this
situation, the angular averaging preserves the gap in the spectrum, approximately given by Eq. (\ref{E_0}).

At $\alpha=45^\circ$, the condition $\Delta_1=\Delta_2$ (which implies the gapless spectrum) is satisfied not only at
$\theta =\pm 45^\circ$ but at any $\theta$ (all trajectories). Then angular averaging does not introduce new qualitative
features (compared to the angle-resolved result), and the DoS at small energies is large according to the results of
Sec.~\ref{sec:Delta1=2}. Comparing with the bulk case, we can say that due to the finite size, the zero-energy peak in
the DoS is smeared but not split. This statement agrees with the results of Nagato and Nagai \cite{NN} who studied this
case by a self-consistent numerical method.

The above results refer to the surface DoS. However, the averaged DoS is linear at low energies also inside the strip,
although the slope is different from Eq. (\ref{nu_av}), because the MGS contributing to this result decay into the bulk
of the sample. For example, in the middle of the strip the angle-averaged DoS differs from Eq. (\ref{nu_av}) by an
additional factor $2\exp(-\Delta d /v_F)$. This result is again valid at $E\ll E_b$; this interval shrinks in the limit
of large $d$, where the DoS is mainly determined by the standard nodal contribution at larger $E$.

\section{Self-consistent numerical results} \label{sec:numerics}

In order to take into account the spatial inhomogeneity of the pair potential, we solve the problem numerically by a
self-consistent numerical method similar to the one used in Ref.~\onlinecite{Amin}. We rewrite the Eilenberger equations
employing the Riccati parametrization,\cite{Schopohl} which is well suited for numerical integration. Starting from the
spatially homogeneous pair potential, we calculate the Green functions over the whole sample, and then use the
self-consistency equation to refine the initial approximation for the pair potential. The procedure is repeated
iteratively until the necessary accuracy is reached.

\begin{figure}
 \includegraphics[width=85mm]{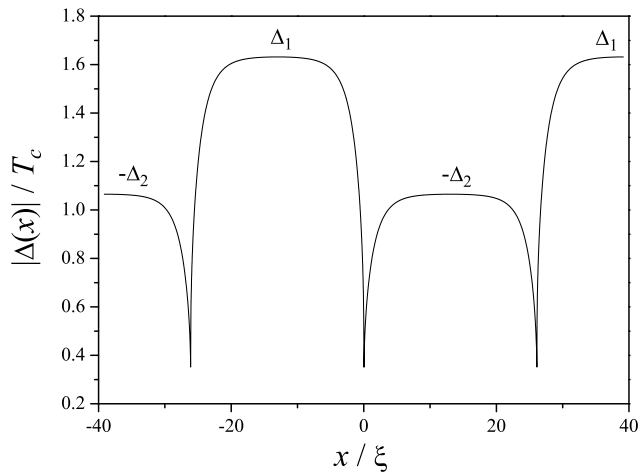}
\caption{Self-consistent profile of the pair potential along the $\theta=40^\circ$ trajectory at the crystalline
orientation $\alpha=20^\circ$. The width of the \textit{d}-wave strip is $L=20\,\xi$, the length of the trajectory
between the successive reflections from the surfaces is $d=L/ \cos 40^\circ$.}
 \label{fig:Delta_sc}
\end{figure}

An example of the pair potential along a trajectory, calculated self-consistently, is shown in Fig.~\ref{fig:Delta_sc}.
In anisotropic superconductors, surfaces lead to pair breaking and suppression of the pair potential. The region where
the suppression takes place, has the characteristic scale of $\xi$, hence at $L \gg \xi$ the piecewise constant $\Delta$
(see Fig.~\ref{fig:Delta12}) is a good approximation.

\begin{figure}
 \includegraphics[width=85mm]{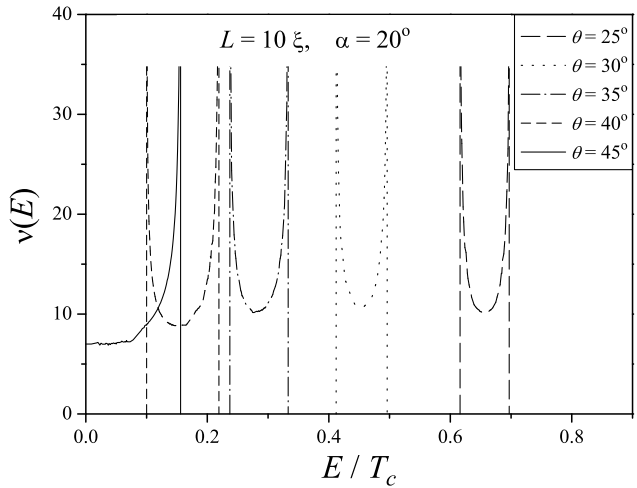}
\caption{The low-energy angle-resolved DoS for several trajectories $\theta$ at the width of the strip $L =10\,\xi$ and
the crystalline orientation $\alpha=20^\circ$.}
 \label{fig:DoS_10}
\end{figure}

\begin{figure}
 \includegraphics[width=85mm]{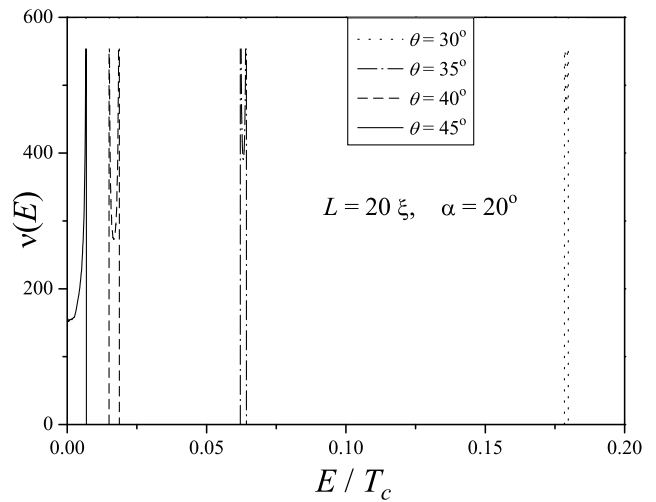}
\caption{The same as in Fig.~\ref{fig:DoS_10} but for $L=20\,\xi$.}
 \label{fig:DoS_20}
\end{figure}

Figures~\ref{fig:DoS_10} and~\ref{fig:DoS_20} demonstrate the angle-resolved DoS for two values of the strip width:
$L=10\,\xi$ and $L=20\,\xi$. These numerical results agree with the analytical theory developed in
Sec.~\ref{sec:analytics}. Indeed, we observe that the spectrum consists of the energy bands, both along the nodal
direction ($\theta=25^\circ$)\cite{nodal} and the directions for which the MGS exist in the limit $L\to\infty$ ($\theta
= 30^\circ$, $35^\circ$, $40^\circ$, and $45^\circ$). The DoS is minimal in the centers of the bands and demonstrates
the square-root singularities on their edges.

The energy gap for the lowest band along the nodal direction ($\theta=25^\circ$) is larger than for the MGS directions;
this agrees with our analytical results according to which the gap for the nodal direction does not contain an
exponentially small factor [see Eq. (\ref{E_0}) and note that the energy gap is of the same order as the center of the
band since the width of the band is much smaller]. The lowest energy band for the nodal direction can be seen in
Fig.~\ref{fig:DoS_10}, while in Fig.~\ref{fig:DoS_20} it is beyond the demonstrated energy range.

The energy bands for the MGS directions are gapped if $\theta\ne 45^\circ$, which corresponds to the case $\Delta_1 \ne
\Delta_2$ studied in Sec.~\ref{sec:Delta1greater2}. The $\theta =45^\circ$ direction corresponds to the case $\Delta_1
=\Delta_2$ studied in Sec.~\ref{sec:Delta1=2}; the energy band is gapless in this case, its center is $E=0$ where the
DoS reaches a minimum.

As $L$ increases, the energy gaps become smaller and the bands become narrower.

The numerical results allow us to directly check the accuracy of our analytical estimates. For example, we consider the
$\theta=40^\circ$ direction at $L=20\,\xi$. Then the edges of the energy band are given by Eqs. (\ref{Emin}) and
(\ref{Emax}) which yield the center of the band $E_0/T_c = 0.015$ and the width of the band $\delta E /T_c = 0.0029$,
while the numerical results are $E_0/T_c = 0.017$ and $\delta E /T_c = 0.0037$ (see Fig.~\ref{fig:DoS_20}). Thus we
conclude that our analytical results provide reasonable quantitative accuracy while capturing all qualitative features
of the DoS.

\section{Conclusions} \label{sec:conclusions}

We have considered the effect of the finite size in the \textit{ab}-plane on the density of states (DoS) in
\textit{d}-wave superconductors. We assumed finiteness in one direction, thus we deal with quasi-two-dimensional strip
instead of quasi-two-dimensional plane. We assumed arbitrary crystalline orientation, described by the angle $\alpha$
between the \textit{a}-axis and the normal to the surfaces.

The problem is solved analytically neglecting the suppression of the pair potential near the surfaces; the results are
confirmed by a self-consistent numerical calculation. In the relevant limits, our results agree with previous studies.

In the bulk, the DoS is gapless along the nodal directions, while the presence of a surface leads to formation of
another type of the low-energy states, the midgap states with zero energy. Due to the finite size of the superconductor,
the spectrum of nodal quasiparticles acquires an energy gap. The midgap states acquire the angle-dependent gap that
vanishes at $\theta=45^\circ$ ($\theta$ is the angle between the trajectory and the normal to the surface); this result
is valid unless the crystal is $0^\circ$- or $45^\circ$-oriented ($\alpha\ne 0^\circ$ or $45^\circ$). In the special
case of $\alpha=0^\circ$, the midgap states are absent, and the spectrum is gapped for all trajectories (including the
nodal directions). On the opposite, in the case of $\alpha=45^\circ$, the spectrum is gapless for all trajectories
$\theta$. In all the cases, the angle-resolved DoS consists of energy bands that are formed similarly to the
Kronig--Penney model.

In the special case of $\alpha=0^\circ$, the angle-averaged DoS has a gap. In the special case of $\alpha=45^\circ$, the
angle-averaged DoS is finite at low energies.

At $\alpha \ne 0^\circ$ or $45^\circ$, the angle-averaged surface DoS is strongly suppressed due to finite size, while
remains gapless and behaves linearly at small energies. The low-energy contribution comes from the trajectories with
$\theta\approx 45^\circ$, hence we can expect that the energy gap survives upon angular averaging if one measures the
\textit{transport} DoS in the case when the $45^\circ$-angle contribution is suppressed by the transparency of the
tunneling barrier.

In our model, we did not take into account such imperfections in the system as bulk
impurities\cite{Hirschfeld,BSM,PBBI,KD,Asano} and interface roughness.\cite{MS,BSB,GK} We expect them to smear the
features discussed in the present paper. In particular, the energy bands (in the angle-resolved DoS) separated from zero
by gaps, exist only when imperfections are weak, otherwise the bands are smeared and finite DoS appears at zero energy.
A rough estimate demonstrates that smearing of the bands due to the bulk impurities exceeds the width of the bands if
the electronic mean free path $l$ is smaller than the width of the strip $L$. However, the effects discussed above can
survive in the case of rare impurities, when $l \gg L$ and long intervals along the strip (between two neighboring
impurities) can be considered clean.

%A rough estimate demonstrates that smearing of the bands due to the bulk impurities becomes comparable to the width of
%the bands already at exponentially large electronic mean free path, $l/ d \sim (\Delta d/v_F)^{-2} \exp(2\Delta d/v_F)$.
%Thus our model corresponds to the ideally clean limit.

Our model is not material-specific and can be in principle applied to any \textit{d}-wave superconductor. However, the
only class of materials where the \textit{d}-wave symmetry is established at present, is the high-$T_c$ oxides, where
the coherence length $\xi$ is very small. Then the finite-size effects discussed above become pronounced at quite small
size $L$.

\begin{acknowledgments}
We are grateful to M.~H.~S.~Amin, Yu.~S.~Barash, A.~M.~Bobkov, I.~V.~Bobkova, M.~V.~Feigel'man, M.~Yu.~Kupriyanov,
G.~Leibovitch, and A.~M.~Zagoskin for helpful discussions. The research was supported by the D-Wave Systems Inc. and the
ESF PiShift program. Ya.V.F. was also supported by the RFBR grants Nos. 03-02-16677 and 04-02-16348, the Russian Science
Support Foundation, the Russian Ministry of Industry, Science and Technology, the program ``Quantum Macrophysics'' of
the Russian Academy of Sciences, CRDF, and the Russian Ministry of Education.
\end{acknowledgments}


\begin{thebibliography}{99}
\bibitem{T_c}
The zero-temperature amplitude of the pair potential in the bulk is related to the critical temperature as $\Delta_0 =
2.14\, T_c$, where the amplitude $\Delta_0$ is defined so that $\Delta(\theta) =\Delta_0 \cos(2\theta-2\alpha)$.
\bibitem{remark}
It is interesting to note that Gallagher solved the Gor'kov equations with subsequent averaging over spatial
oscillations on the scale of $1/k_F$ ($k_F$ is the Fermi wavevector), while we start from the quasiclassical Eilenberger
equations. Thus the correspondence between our results implies that the quasiclassical approximation in the problem is
valid provided $k_F L \gg 1$.
\bibitem{nodal}
At $\alpha=20^\circ$, one of the nodal directions in the bulk is $\theta=-25^\circ$. In the strip geometry, we call the
$\theta=25^\circ$ direction nodal because the corresponding trajectory includes the nodal $\theta=-25^\circ$ segments
(see Fig.~\ref{fig:system}).
\end{thebibliography}
\end{document}